\documentstyle[preprint,prb,aps]{revtex}
\tighten
 
\begin{document}
\draft
\title{Quantum fluctuations in the incommensurate phase of 
CsCuCl$_3$ under transverse magnetic field}
 
\author{Tetsuro Nikuni\cite{Tetsuro}}
\address{Department of Physics, Tokyo Institute of Technology, 
Oh--okayama, Tokyo, 152 Japan}
\address{and Department of Physics, University of Toronto,
Toronto, Ontario, Canada M5S 1A7}
\author{A. E. Jacobs\cite{Allan}}
\address{Department of Physics, University of Toronto,
Toronto, Ontario, Canada M5S~1A7}
 
\date{\today}

\maketitle
 
\begin{abstract}

In low magnetic field, the stacked, triangular antiferromagnet CsCuCl$_3$ 
has a helical structure incommensurate (IC) in the chain direction. 
The IC wavenumber (from neutron--diffraction experiments) decreases with 
increasing field transverse to the chains, as predicted by classical theory, 
but then it has a plateau almost certainly caused by quantum fluctuations. 
Linear spin--wave theory fails because fluctuations have particularly large 
effects in the IC phase. 
An innovative phenomenological treatment of quantum fluctuations yields 
a plateau at approximately the observed value and the observed fields; 
it predicts a transition to the commensurate phase so far not observed. 
Results depend sensitively on a weak anisotropy. 
\end{abstract}
 
 
\section{INTRODUCTION}
 
Compounds of the $ABX_3$ family 
($A=$ Rb, Cs; $B=$ Mn, Fe, Co, Ni, Cu, V; $X=$ Cl,~Br,~I) figure 
prominently in the study of phase transitions in low--dimensional systems. 
Much of the interest in the magnetic--field behavior of 
compounds like CsCuCl$_3$ arises because they are physical realizations 
of models related to the triangular antiferromagnet (TAFM). 
The TAFM ground state is both continuously and discretely (two--fold) 
degenerate, even in the presence of a magnetic field {\bf H} (with magnitude 
$H$ less than the saturation field $H_{\rm S}$), unlike that of the 
square--lattice AFM. 
In a field, 
thermal fluctuations\cite{lee84,kawa84,kawamiya85,korsh86} in classical 
TAFM models and quantum fluctuations\cite{nishi86,chub91} break the 
continuous degeneracy (which is nontrivial because it is not due to a 
symmetry of the Hamiltonian) in the same way, both selecting for example 
the colinear structure at $H\approx H_{\rm S}/3\,$. 

The magnetic properties of CsCuCl$_3$ (with a N\'eel 
temperature\cite{achiwa69} $T_{\rm N}=10.7\,$K) arise from the Cu$^{++}$ 
ions; to a good approximation, these form a triangular lattice of parallel 
chains, the other ions serving to define the structure. 
The major interactions, all nearest--neighbor, are a ferromagnetic exchange 
interaction in the chain or $c$ direction, a weaker antiferromagnetic exchange 
interaction between chains (within the $a$--$b$ planes), and a 
Dzyaloshinskii--Moriya\cite{DM} (DM) interaction also in the $c$ direction. 
Both exchange interactions are nearly isotropic; the latter is frustrated. 
Recent studies\cite{koiso96,ogihara95} of the structure and of the phase 
transition giving rise to the DM term cite earlier literature on these 
topics. 
In the simplified structural model, the classical, zero--temperature, 
zero--field structure is a three--sublattice, $\pm120^\circ$ TAFM structure 
in each $a$--$b$ plane; the spins lie in the planes and rotate from plane 
to plane, forming an incommensurate (IC) helical structure\cite{adachi80}. 

Fluctuation effects are likely large in CsCuCl$_3$ for several reasons: 
the Cu spin is small ($S=1/2$), 
the system is almost one--dimensional (the intrachain interaction 
is much larger than the interchain interaction), 
the interchain interaction is frustrated, 
the exchange interactions are nearly isotropic, 
and the structure is incommensurate. 
Experiments in magnetic field (difficult because $H_{\rm S}=30\,$T) 
indeed find major effects due to quantum fluctuations. 

CsCuCl$_3$ in a longitudinal field (${\bf H}\parallel{\bf c}$) appears to be 
well understood at low temperatures $T$, but the transition at $T_{\rm N}\,$ 
has puzzling features\cite{weber97}. 
The discontinuity\cite{moto78,fed85,nojiri88} in the low--$T$ 
magnetization at $H\approx0.4\,H_{\rm S}$ was shown by Nikuni and 
Shiba\cite{shiba93,nikuni93} to be a novel, fluctuation--induced phase 
transition from the umbrella structure (optimal at small $H$ due to a 
small, easy--plane anisotropy\cite{tanaka92} in the intrachain exchange) to 
a coplanar structure (optimal at larger $H$ due to quantum fluctuations). 
Further experiments\cite{ohta93,chiba94,mino94,ohta94,chiba95}, including 
neutron--diffraction\cite{schotte94} and specific--heat 
measurements\cite{weber97} near $T_{\rm N}\,$, confirmed their analysis. 

Properties for a transverse field (${\bf H}\perp{\bf c}$) are not well 
established. 
Small fields deform the helix, increasing its period; 
in agreement with experiment\cite{mino94,schotte94}, classical (mean--field) 
theory\cite{jacobs93,tomo95} predicts that the IC wavenumber $q$ decreases 
quadratically as $H$ increases and that the curvature increases with $T$. 
The structure remains incommensurate\cite{mino94} up to 
$H\approx0.44\,H_{\rm S}$; it is unknown at larger fields, where classical 
theory\cite{jacobs93,tomo95} predicts an intermediate commensurate 
(C) phase in which each plane has the same three--sublattice structure. 
Classical theory fails at intermediate fields: near $H_{\rm S}/3\,$, 
plateaus are observed in the magnetization\cite{nojiri88} $m$, the 
$^{133}$Cs NMR shift\cite{chiba95} and the wavenumber\cite{mino94}; 
the ESR measurements\cite{ohta94} are not easily interpreted. 
The plateau in $m$, as for the TAFM\cite{nishi86,chub91}, is due to quantum 
fluctuations\cite{jacobs93} (the analysis was done for the C state, but the 
result for the IC state should not differ much). 
Quantum fluctuations are surely responsible also for the plateau in $q$, 
but this remains to be demonstrated. 
Recent specific--heat, magnetization and neutron--diffraction 
measurements\cite{werner97} near $T_{\rm N}$ suggest major effects due to 
thermal fluctuations: $T_{\rm N}$ increases with field (as in the 
TAFM\cite{lee84,kawamiya85,miya86JPSJ}) and a new phase appears. 

The following examines quantum fluctuations in the incommensurate phase of 
CsCuCl$_3\,$. 
Section II.A describes the Hamiltonian. 
Section II.B describes classical results (from solution of the 
Euler--Lagrange equations); 
surprisingly, the spins do not remain in the $a$--$b$ planes at 
intermediate fields (as found also by Jensen\cite{jensen}). 
Section II.C describes a linear spin--wave (LSW) analysis based on the 
classical results; 
this fails because LSW theory does not find the C--state and IC--state 
energies to equivalent accuracy. 
Section III introduces a phenomenological treatment of quantum fluctuations 
and shows that it works well for the C state; 
the same approach applied to the IC state yields a plateau in the 
wavenumber at approximately the observed value and at approximately the 
observed fields. 
Although results are rather sensitive to an anisotropy parameter, 
we conclude that a commensurate phase should appear at a 
field well below the saturation field, likely below $0.5\,H_{\rm S}\,$. 

\section{Hamiltonian and analysis}

\subsection{Hamiltonian} 

The Hamiltonian corresponding to the simplified structure is 
$${\cal H}=\sum_{in}\Bigl[\,-\,2\,J_0\,{\bf S}_{in}\cdot{\bf S}_{i,n+1}
                     +2\,\eta\,J_0\,S^{(z)}_{in} S^{(z)}_{i,n+1}
-D\,\hat{\bf z}\cdot({\bf S}_{in}\times{\bf S}_{i,n+1})$$
$$+\,J_1\mathop{{\sum}'}_k{\bf S}_{in}\cdot{\bf S}_{kn}
-g\,\mu_{\rm B}\,H\,\hat{{\bf x}}\cdot{\bf S}_{in}\,\Bigr]\ ,\eqno(1)$$
where ${\bf S}_{in}$ is the spin operator at the $i$--th site in the 
$n$--th $a$--$b$ plane, $\hat{\bf z}$ and $\hat{\bf x}$ are unit vectors in 
the $c$ and $a$ directions, and the $k$ sum is over the six, in--plane, 
nearest neighbors of the site $in$. 
The first term ($\propto J_0$) is the isotropic, ferromagnetic 
exchange interaction between spins in nearest--neighbor planes, 
the second is an anisotropic correction (of easy--plane type) to the first, 
the third ($\propto D$) is the interplane DM interaction, 
the fourth ($\propto J_1$) is the isotropic, antiferromagnetic exchange 
interaction between nearest-neighbor spins in the $a$--$b$ planes, and the 
fifth is the Zeeman energy in a transverse field ${\bf H}$. 
The values of the coefficients have been estimated in several 
articles\cite{achiwa69,hyodo81,tazuke81,tanaka92,mekata95}; 
we use $J_0=28\,$K, $\eta J_0=0.24\,$K, $J_1=4.9\,$K and $D=5\,$K. 
We omit the dipole--dipole interaction (which can induce an IC state modulated 
in the planes\cite{shiba82}), the anisotropy of the interchain interaction, 
and several small, related effects (namely the displacement of the Cu ions 
from the $c$ axis, the component of the DM vector perpendicular to the $c$ 
axis, and the $z$ component of the magnetizations\cite{schotte94,mekata95}). 
The saturation field $H_{\rm S}=30\,$T, above which each spin is aligned 
with the field, is $18J_1S/(g\mu_{\rm B})$ in terms of the model; 
the reduced field is $h=H/H_{\rm S}\,$. 
A longitudinal field $({\bf H}\parallel{\bf c})$ maintains the axial symmetry 
of zero field, and the DM term can be eliminated\cite{shiba93,nikuni93}. 
A transverse field $({\bf H}\perp{\bf c})$ breaks the symmetry, and the 
DM term comes into full play. 

The intrachain exchange term $(J_0$) favors a state with spins parallel in 
adjacent layers while the smaller DM term favors a rotation by $\pi/2$ per 
layer; 
the result at low fields is a helical structure, incommensurate in the 
$c$ direction. 
The IC wavenumber $q$ is small ($\approx2\pi/71$) at $h=0$ where 
the continuous degeneracy of the classical TAFM ground state corresponds 
to a mere shift in the origin of the coordinate system; 
the classical degeneracies remain for a transverse field $<H_{\rm S}\,$, 
even if the spins are confined to the planes. 

Both the DM term (inactive for the possible C state in transverse field) and 
the much smaller anisotropy term\cite{tanaka92} (responsible for the phase 
transition in longitudinal field\cite{shiba93,nikuni93}) favor spins in the 
$a$--$b$ planes. 
For the IC state in transverse field, it is natural to assume that the DM term 
confines the spins to the planes at all fields (as it does at small $H$), 
and to ignore the $\eta$ term. 
Surprisingly, in--plane spins are unstable at intermediate fields for 
isotropic exchange (as found also by Jensen\cite{jensen}); the instability 
persists, but is much reduced, for the experimental anisotropy. 

We assume that the three--sublattice structure is 
maintained\cite{mino94,schotte94} at all fields $<H_{\rm S}\,$. 
A continuum approximation\cite{jacobs93} for the $c$--direction dependence is 
valid, but we require the discrete model because we approximate IC states 
as high--order commensurate states; 
in principle, we miss some features of IC states, but on the other hand 
we find that the states are not pinned. 
We use periodic boundary conditions ${\bf S}_{j,l+L}={\bf S}_{jl}\,$, where 
$j=1,2,3$ is the sublattice index and $l=1,\cdots,L$ is the layer index. 

\subsection{Classical analysis}

In the classical approximation, the spin operators ${\bf S}_{in}$ become 
classical vectors of length 1/2, and the Hamiltonian of Eq.(1) becomes 
the energy function $E_{\rm cl}(\{{\bf S}_{in}\})$. 
With Lagrange multipliers $\lambda_{jl}$ for the constraints 
${\bf S}_{jl}\cdot{\bf S}_{jl}-S^2=0$, the Euler--Lagrange equations are 
$$-\,2\,J_0\,({\bf S}_{j,l-1}+{\bf S}_{j,l+1})
+2\,\eta\,J_0\,(S^{(z)}_{j,l-1}+S^{(z)}_{j,l+1})\ \hat{\bf z}
-D\,({\bf S}_{j,l+1} -{\bf S}_{j,l-1})\times\hat{\bf z} $$
$$+\,6\,J_1\,({\bf S}_{j-1,l}+{\bf S}_{j+1,l})
-18\,J_1\,S\,h\,\hat{\bf x}-2\,\lambda_{jl}\,{\bf S}_{jl}=0\ ;\eqno(2)$$ 
the solutions are not pinned and so a condition such as 
$\hat{\bf y}\cdot{\bf S}_{11}=0$ is necessary. 
Ref.\onlinecite{jacobs93} studied the continuum version of Eq.(2) (without 
the anisotropy term) for in--plane spins. 

The Euler--Lagrange equations have many solutions. 
The one optimal at a given field is too often neither simple nor obvious, 
and so it was necessary to generate scores of solutions and follow them as 
functions of the field (in steps $\Delta h=0.01$), minimizing the energy 
of each with respect to $L\,$; 
composite solutions (with increased winding number) gave greater accuracy 
when such was desired. 
Random starting configurations were generated at various values of $h$ and 
$L$, relaxed by conjugate--gradient minimization of the energy, and 
then used to start solution of the Euler--Lagrange equations. 

Classical theory cannot explain the plateau in $q\,$, but it is a 
necessary preliminary to the LSW analysis and it sheds light on two questions: 
(1) does a commensurate state intervene between the incommensurate and 
aligned states, and (2) do the spins lie in the $a$--$b$ planes at all fields? 
The answers depend crucially on the size of the anisotropy parameter $\eta$. 

For isotropic exchange ($\eta=0$), the optimal structure is incommensurate up 
to $H_{\rm S}\,$. 
For $h\leq0.38\,$, there are many solutions (as in the continuum 
approximation\cite{jacobs93}), but they are well separated in energy, and the 
spins remain in the planes for the optimal solution (the 111 solution of 
Ref.\onlinecite{jacobs93}), which evolves continuously from the solution at 
$h=0\,$. 
For $h\agt0.38\,$, the spins break out of the planes (in the optimal solution); 
this is surprising, for each intrachain term in the energy is optimized or 
neutral if the spins lie in the planes, and so is the sum of the interchain 
and field terms. 
In contrast, we verified that the spins remain in the planes for all fields 
in the case of a model\cite{moriya82} with a ferromagnetic chain--chain 
interaction. 
The IC phase for $h\agt0.38$ is, however, very complicated: 
the solutions are far more numerous, they differ only slightly in energy for 
the most part, and the low--energy solutions are not the simplest; 
ground--state energy crossings make it impossible to identify with 
confidence the optimal solution in most of this range. 
In short, the IC phase is glassy for fields in much of the range 
$0.38\alt h<1$, the ground state changing often and unpredictably with 
field, from one complicated configuration to another. 

The weak, easy--plane anisotropy\cite{tanaka92} $\eta=8.6\times10^{-3}$ 
changes the phase diagram substantially, opening a large window for the C 
phase and destroying the glassy phase. 
For $0\leq h\leq0.41\,$, the in--plane IC solution of the previous paragraph 
is optimal; it extends to larger $h$ because of the anisotropy. 
For $0.42\leq h\leq0.50\,$, a simple, out--of--plane, IC solution is optimal 
(but other IC solutions are close in energy); 
that is, both the DM and anisotropy terms acting in concert fail to confine 
the spins to the planes at all fields. 
For $0.51\leq h<1\,$, the commensurate (C) state is optimal. 
All transitions are second--order. 
If $\eta$ were only slightly larger (say a few \%), then likely the spins 
would remain in the planes at all fields, and the IC phase make a 
second--order transition to the C phase\cite{jacobs93} at $h\approx0.47\,$. 

\subsection{Linear spin--wave analysis} 

The following describes a linear--spin--wave (LSW) analysis of quantum 
fluctuations in incommensurate (IC) states. 
To our knowledge, fluctuations have previously been studied only in IC 
states which are sinusoidal or nearly so. 
The analysis was restricted to in--plane spins and isotropic exchange 
($\eta=0$), and so results were obtained only for $h\leq0.38\,$. 

A Holstein--Primakoff transformation\cite{HP40} to boson operators (using a 
local coordinate system\cite{jacobs93}) and an expansion about the 
classical solutions give the Hamiltonian as 
$${\cal H}=E_{\rm cl}+{\cal H}_1+{\cal H}_2+{\rm O}(\sqrt{S})\ ;\eqno(3)$$ 
the classical spins are not colinear, and so the expansion parameter is 
$1/\sqrt{S}\,$ (rather than $1/S$), as in Refs.\onlinecite{jacobs93} 
and\onlinecite{chub94}. 
The classical angles $\phi_{jl}$ between the spins and the field are 
determined by minimizing the classical energy $E_{\rm cl}$ (of order $S^2$) 
of the $N$ spins: 
$$E_{\rm cl}=S^2\,\frac{N}{3L} \sum_{j=1}^3 \sum_{l=1}^L
\Bigl[-\,2\,J_0\cos(\phi_{j,l+1}\!-\!\phi_{jl}) 
             -D\sin(\phi_{j,l+1}\!-\!\phi_{jl})$$
$$\qquad\qquad\qquad+\,6\,J_1\cos(\phi_{j+1,l}\!-\!\phi_{jl}) 
-18\,J_1\,h\cos\phi_{jl} \ \Bigr]\ .\eqno(4)$$
The term ${\cal H}_1$ (of order $S^{3/2}$) is linear in the boson 
operators; 
it vanishes for the classical angles $\phi_{jl}$ and so is omitted. 
The LSW Hamiltonian ${\cal H}_2$ (of order $S$) has the standard form 
(quadratic in the boson operators): 
$${\cal H}_2=
-\frac{S}{2}\frac{N}{3L} \sum_{jl} C_{jl}
+\frac{S}{2}\sum_{\bf k}\sum_{jl}\sum_{j^\prime l^\prime}
\Bigl\{ {\cal A}_{jl,j^\prime l^\prime}({\bf k})\,
 \left[b_{jl}^\dagger({\bf k})
            \,b_{j^\prime l^\prime}^{\phantom\dagger}({\bf k})
 +b_{jl}^{\phantom\dagger}({\bf-k})
            \,b_{j^\prime l^\prime}^\dagger({\bf-k})\right]$$
$$\qquad\qquad\qquad\qquad\qquad\qquad\qquad  
    +{\cal B}_{jl,j^\prime l^\prime}({\bf k})\,
      \left[b_{jl}^\dagger({\bf k})
            \,b_{j^\prime l^\prime}^\dagger({\bf-k})
           +b_{jl}^{\phantom\dagger}({\bf-k})
            \,b_{j^\prime l^\prime}^{\phantom\dagger}({\bf k})\right]
						\Bigr\}\ .\eqno(5)$$
The ${\bf k}$ sum runs over the $N/(3L)$ points in the first Brillouin 
zone; the $3L\times3L$ Hermitian matrices ${\cal A}$, ${\cal B}$ and 
${\cal C}$, the last diagonal (${\cal C}_{jl,j^\prime l^\prime} 
=C_{jl}\,\delta_{jj^\prime}\,\delta_{ll^\prime})$ 
and independent of ${\bf k}$, are 
$$\left[{\cal A}({\bf k}) +{\cal B}({\bf k}) -{\cal C} 
          \right]_{jl,j^\prime l^\prime}=
  -2J_0\cos(\phi_{jl^\prime}\!-\!\phi_{jl})\,\delta_{j^\prime j}
      \left(e^{ ik_z}\,\delta_{l^\prime,l+1}
           +e^{-ik_z}\,\delta_{l^\prime,l-1}\right)  $$
$$   -D\sin(\phi_{jl^\prime}\!-\!\phi_{jl})\,\delta_{j^\prime j}
      \left(e^{ ik_z}\,\delta_{l^\prime,l+1}
           -e^{-ik_z}\,\delta_{l^\prime,l-1}\right)  $$
$$ +6J_1\cos(\phi_{j^\prime l}\!-\!\phi_{jl})\,\delta_{l^\prime l}
     \left(\nu_{\bf k}^{\phantom\ast}\,\delta_{j^\prime,j+1}
          +\nu_{\bf k}^\ast          \,\delta_{j^\prime,j-1}\right)
\ ,\eqno(6)$$
$$\left[{\cal A}({\bf k}) -{\cal B}({\bf k}) -{\cal C} 
          \right]_{jl,j^\prime l^\prime}=
 -2J_0\delta_{j^\prime j}\left(e^{ ik_z}\delta_{l^\prime,l+1}
                              +e^{-ik_z}\delta_{l^\prime,l-1}\right)
 +6J_1\delta_{l^\prime l}
     \left(\nu_{\bf k}^{\phantom\ast}\delta_{j^\prime,j+1}
          +\nu_{\bf k}^\ast          \delta_{j^\prime,j-1}\right)
\ ,\eqno(7)$$ 
$$C_{jl}=
2\,J_0\,\left[\cos(\phi_{j,l+1}\!-\!\phi_{jl})
             +\cos(\phi_{jl}   \!-\!\phi_{j,l-1})\right]
    +D\,\left[\sin(\phi_{j,l+1}\!-\!\phi_{jl})
             +\sin(\phi_{jl}   \!-\!\phi_{j,l-1})\right]$$
$$-6\,J_1\,\left[\cos(\phi_{j+1,l}\!-\!\phi_{jl})
                +\cos(\phi_{j-1,l}\!-\!\phi_{jl})\right]
+\,18\,J_1\,h\cos\phi_{jl} \ ,\eqno(8)$$
where $\nu_{\bf k}$ is the in--plane structure factor 
$$\nu_{\bf k}= {\textstyle{1\over3}}\,\left\{\exp(ik_x)
+\exp\left[{\textstyle{1\over2}}\,i\left(-k_x\!+\!\sqrt{3}k_y\right)\right] 
+\exp\left[{\textstyle{1\over2}}\,i\left(-k_x\!-\!\sqrt{3}k_y\right)\right] 
\right\}\ .\eqno(9)$$
After a standard transformation\cite{HP40,bogo47} to creation and annihilation 
operators $\gamma^\dagger_{jl}$ and $\gamma^{\phantom\dagger}_{jl}$ for the 
spin--wave excitations, ${\cal H}_2$ takes the form 
$${\cal H}_2=
-{S\over2}{N\over{3L}}\sum_{jl} C_{jl}
+S\sum_{{\bf k}}\sum_{jl}\epsilon_{jl}({\bf k})
\left[\gamma_{jl}^\dagger({\bf k})\,\gamma_{jl}^{\phantom\dagger}({\bf k})
+{1\over2}\right]\ .\eqno(10)$$
The ground--state value is $E_2=\langle0|{\cal H}_2|0\rangle$; 
the excitation energies $\epsilon$ ($\geq0$) are found from 
$$\det\,\left[\,\left({\cal A}-{\cal B}\right)
 \left({\cal A}+{\cal B}\right)-\epsilon^2\,\hat1\,\right]=0\ .\eqno(11)$$
The eigenvalue problem for $\epsilon_{jl}^2({\bf k})$ is Hermitian, but the 
result for $\epsilon_{jl}({\bf k})$ can be imaginary (as when the 
classical, in--plane solutions become unstable at intermediate fields). 

The LSW analysis of quantum fluctuations in the C state is 
straightforward\cite{jacobs93}; 
of course the antisymmetric DM term does not appear, and so 
quantum selection occurs just as in the TAFM\cite{chub91}.
For the IC state (the 111 state of Ref.\onlinecite{jacobs93}), the total 
energy $E_{\rm cl}+E_2$ was found as a function of the period $L$ and 
minimized with respect to $L$ to give the optimal wavenumber $q=2\pi/L$. 
In the relevant field region, the optimal $L$ ranged from $\approx70$ 
$a$--$b$ plane spacings to $\approx150$, large enough to justify our 
treatment of the IC state as a high--order commensurate state and 
small enough that the diagonalization was practicable. 

The IC wavenumber (Figure 1) may flatten out with field, but the 
assumption of in--plane spins fails for $h\agt0.38\,$. 
More seriously, LSW theory gives a transition to the C state at lower field, 
$h\approx0.32\,$. 
The result for $\eta=8.6\times10^{-3}$ would not be qualitatively different. 

LSW theory fails because it finds the C--state energy more accurately than the 
IC--state energy. 
The C--state energy is found by choosing the classical configuration (for 
example, the colinear state at $h=1/3$) which minimizes the total energy 
$E_{\rm cl}+E_2$. 
Such a choice is not possible for the IC state and so LSW theory does not 
take into account sufficiently the breaking of the continuous degeneracy in 
determining the spin structure; 
moreover, the classical IC state is a poor approximation to the quantum 
IC state, as we now discuss.
Much of the classical IC structure\cite{jacobs93} can be described as a 
sequence of degenerate commensurate states (a ``spatially varying 
commensurate state''); 
domain walls (where the spins assume orientations classically forbidden in 
the TAFM) are well defined only very close to the IC--C transition. 
But quantum fluctuations destroy the continuous degeneracy and so the IC 
state at even small fields will have the conventional structure, namely 
distinct regions where the order parameter is nearly constant at a 
commensurate value, separated by walls where it varies rapidly. 
This structure is very different from the classical structure, and thus is 
difficult to obtain by the perturbative approach of the $1/S$ expansion.

Extension of the analysis to higher order (including spin rearrangement due 
to quantum fluctuations) would also not find the two energies to the same 
accuracy; and it is 
computationally out of reach, entirely without appeal, and even not possible 
if $\eta=0$ (the classical ground state cannot be found for $h\agt0.38$). 
A very different approach is required. 

\section{Phenomenological treatment of quantum fluctuations}

The most important effect of quantum fluctuations in this problem is the 
breaking of the TAFM classical degeneracy. 
The following describes an innovative phenomenological method to treat 
quantum fluctuations in the TAFM part of the Hamiltonian. 
We represent the quantum part of the TAFM energy (terms proportional to 
$S^1$, $S^0$, {\it etc.}) by the term 
$$ E_{\rm fluct}=-\frac{J_2}{S^2} \sum_{\langle ij\rangle}\sum_n
\left({\bf S}_{in}\cdot{\bf S}_{jn}\right)^2\eqno(12{\rm a})$$
with $J_2>0$; for in--plane spins, 
$$E_{\rm fluct}=-J_2\,S^2\frac{N}{L}\sum_{j=1}^3\sum_{l=1}^L
\cos^2(\phi_{j+1,l}\!-\!\phi_{jl}) \ . \eqno(12{\rm b})$$
The motivation for this form of $E_{\rm fluct}$ is partly that it favors the 
colinear state. 
A biquadratic coupling term is important in the classical theory\cite{saka84} 
of the plateau\cite{sue81} in the magnetization of the $S=7/2$ system 
C$_6$Eu$\,$, and Ref.\onlinecite{chub91} pointed out that such a term in the 
Hamiltonian (with $S>1/2$) may have the same consequences as quantum 
fluctuations. 
Our treatment is phenomenological because a biquadratic coupling term 
is not allowed in the Hamiltonian for CsCuCl$_3$ (because $S=1/2$), and 
therefore the term appears here in a fundamentally different way than 
previously.  

To test whether the phenomenological term describes qualitatively the 
effects of quantum fluctuations, we applied the formalism to the TAFM, 
minimizing the total energy 
$$E=N\sum_{j=1}^3\left[\,2\,J_1\,{\bf S}_j\cdot{\bf S}_{j+1}
   -g\,\mu_{\rm B}\,{\bf H}\cdot{\bf S}_j/3
-J_2\left({\bf S}_j\cdot{\bf S}_{j+1}\right)^2/S^2\,\right]\eqno(13)$$ 
(this is just $E_{\rm cl}+E_{\rm fluct}$) for the five TAFM states shown in 
Figure 2; the direction of {\bf H} is irrelevant and all states have the 
same classical energy. 

\noindent (a) The umbrella state: 
The angle $\phi$ between the spins and the field is found from 
$$18\,J_1\,S\cos\phi-9\,J_2\,S\cos\phi\,(3\cos^2\phi-1)
=g\,\mu_{\rm B}\,H\ .\eqno(14)$$ 

\noindent (b) The coplanar state $\phi_1=\pi$, $\phi_3=-\phi_2$ with 
$\sin\phi_2\neq0$: The angle $\phi_2$ is found from 
$$6\,J_1\,S\,(2\cos\phi_2-1)-6\,J_2\,S\cos\phi_2\,(1+2\cos2\phi_2)
=g\,\mu_{\rm B}\,H\ .\eqno(15)$$

\noindent (c) The colinear state $\phi_1=\pi$, $\phi_2=\phi_3=0$. 

\noindent (d) The coplanar state $\phi_3=\phi_2$ with $\sin\phi_2\neq0$: 
The angles $\phi_1$ and $\phi_2$ are determined by 
$$12\,J_1\,S\sin(\phi_1\!-\!\phi_2)-6\,J_2\,S\sin(2\phi_1\!-\!2\phi_2)
=g\,\mu_{\rm B}\,H \sin\phi_1\ ,\eqno(16{\rm a})$$
$$\sin\phi_1+2\sin\phi_2=0\ .\eqno(16{\rm b})$$

\noindent (e) The coplanar state $\phi_1=0$, $\phi_3=-\phi_2$ with 
$\sin\phi_2\neq0$: The angle $\phi_2$ is found from 
$$6\,J_1\,S\,(2\cos\phi_2+1)-6\,J_2\,S\cos\phi_2\,(1+2\cos2\phi_2)
=g\,\mu_{\rm B}\,H\ .\eqno(17)$$

\noindent State (b) is optimal at low fields, 
$0<H<H_1=6(J_1\!-\!3J_2)S/g\mu_{\rm B}\,$, the region over which it exists. 
The colinear state (c) is optimal at intermediate fields, 
$H_1<H<H_2=6(J_1\!+\!J_2)S/g\mu_{\rm B}\,$; 
the magnetization in this region is one--third the saturation value.
State (d) is optimal at higher fields, $H_2<H<H_{\rm S}=
18(J_1\!-\!J_2)S/g\mu_{\rm B}\,$, the region over which it exists. 
The aligned state $\phi_j=0$ is optimal for $H>H_{\rm S}\,$. 
Except for the definitions of the fields $H_1\,$, $H_2$ and $H_{\rm S}\,$, 
these are just the $1/S$ results\cite{chub91} for quantum fluctuations: 
the colinear state is stable in a region about $h=1/3\,$, states (a) and 
(e) are optimal at no field $H$ $>0$ and $<H_{\rm S}\,$, {\it etc}. 
However, quantum fluctuations do not renormalize\cite{chub91} $H_{\rm S}\,$, 
and the expressions for the fields $H_1$ and $H_2$ differ qualitatively. 

Having established that the term $E_{\rm fluct}$ provides a reasonable 
description of quantum fluctuations in the TAFM, we next compare it 
quantitatively with the LSW energy $E_2$ for the five commensurate states 
of CsCuCl$_3$ derived from the TAFM states of Figure 2. 
The energies $E_2$ were found by standard methods\cite{jacobs93} with 
parameter values $J_0=28\,$K, $\eta=0$, $J_1=4.9\,$K and $S=1/2\,$. 
The energies $E_{\rm fluct}\,$, on the other hand, are the same as for the 
TAFM and are independent of $J_0\,$. 
To make a fair comparison ($E_2$ is only the leading quantum correction), we 
found $E_{\rm fluct}$ to only first order in $J_2\,$; this means, for 
example, that $\cos\phi=h=g\mu_{\rm B}H/18J_1S$ for the umbrella state. 
Figure 3 plots the two energies for the first four states relative to 
state (e) (which exists over the entire field range $0\leq H<H_{\rm S}$ 
and so is a convenient reference state). 
Good agreement is obtained for $J_2=0.2\,$K and so we expect that 
the phenomenological term $E_{\rm fluct}$ captures well enough 
the breaking of the classical degeneracy by quantum fluctuations.
Because $E_2$ is only the first--order correction, a moderate 
adjustment of $J_2$ is acceptable; 
an effective $J_2$ smaller than $0.2\,$K is suggested by the next 
correction\cite{nikuni95} at fields just below $H_{\rm S}\,$. 
A separate comparison should really be made for $\eta=8.6\times10^{-3}$ but 
this seems unwarranted. 

The phenomenological approach was then applied to the IC phase of CsCuCl$_3$ 
in transverse field. The total energy is $E_{\rm cl}+E_{\rm fluct}\,$. 
The Euler--Lagrange equations are 
$$-\,2\,J_0\,({\bf S}_{j,l-1}+{\bf S}_{j,l+1})
+2\,\eta\,J_0\,(S^{(z)}_{j,l-1}+S^{(z)}_{j,l+1})\ \hat{\bf z}
-D\,({\bf S}_{j,l+1} -{\bf S}_{j,l-1})\times\hat{\bf z} 
+\,6\,J_1\,({\bf S}_{j-1,l}+{\bf S}_{j+1,l})$$
$$-g\,\mu_{\rm B}\,H\,\hat{\bf x} -2\,\lambda_{jl}\,{\bf S}_{jl}
-\,6\,J_2\,S^{-2}
\left[({\bf S}_{j-1,l} \cdot {\bf S}_{jl})\,{\bf S}_{j-1,l}
     +({\bf S}_{j+1,l} \cdot {\bf S}_{jl})\,{\bf S}_{j+1,l}\right]
=0\ ;\eqno(18{\rm a})$$ 
for in--plane spins, these simplify to 
$$-\,2\,J_0\,S\left[\sin(\phi_{jl}\!-\!\phi_{j,l-1})
                   +\sin(\phi_{jl}\!-\!\phi_{j,l+1})\right]
  -D       \,S\left[\cos(\phi_{jl}\!-\!\phi_{j,l+1})
                   -\cos(\phi_{jl}\!-\!\phi_{j,l-1})\right]$$
$$+\,6\,J_1\,S\left[\sin(\phi_{jl}\!-\!\phi_{j-1,l})
                   +\sin(\phi_{jl}\!-\!\phi_{j+1,l})\right]
  -g\,\mu_{\rm B}\,H\sin\phi_{jl}$$
$$-\,3\,J_2\,S\left[\sin(2\phi_{jl}\!-\!2\phi_{j-1,l})
                   +\sin(2\phi_{jl}\!-\!2\phi_{j+1,l})\right]=0\ .
\eqno(18{\rm b})$$
As described in Section II.B, scores of solutions of these 
equations were generated and followed in field, in steps 
$\Delta h=0.01$ [with $h=g\mu_{\rm B}/(18J_1S)$]. 
In the following, we use the value $J_2=0.13\,$K which gives a plateau in 
$q\,$; there is no plateau for significantly smaller values, and $q$ is 
nonmonotonic for significantly larger values ($J_2=0.17\,$K for 
example)\cite{thesis}. 
The phenomenological term does not change the zero--field spin configuration, 
a simple helical structure with wavenumber $q_0=\arctan(D/2J_0)\,$; 
the result for weak fields is almost the same as for $J_2=0$. 
But quantum fluctuations, represented phenomenologically, make major 
differences for $H\agt H_{\rm S}/3\,$.

For isotropic exchange ($\eta=0$), the ground state is incommensurate up to 
$H_{\rm S}\,$. 
The spins lie in the planes until $h=0.38\,$; 
above this field, two out--of--plane solutions compete, their energies 
crossing several times; quantum fluctuations destroy the glassy phase.  

For $\eta=8.6\times10^{-3}$, the low--field, in--plane solution is optimal 
for $h\leq0.42\,$; 
as for the classical analysis (Section II.B), the weak, easy--plane anisotropy 
opens a large window for the commensurate state, which is optimal for 
$h\geq0.44$ up to $H=H_{\rm S}\,$; 
an out--of--plane IC solution is optimal over such a small field 
range ($\Delta h<0.013$ about $h=0.43$) that we ignore it. 
Figure 4 compares the experimental results (available only to $H=13\,$T 
or $h=0.43$) for the IC wavenumber with the theoretical values. 
The agreement with experiment is moderately good. 
The plateau occurs at about the observed value of $q/q_0$ and over 
about the correct field range; 
of course the value $J_2=0.13\,$K was selected to give a plateau, but 
the position of the plateau is not adjustable. 
The major difference is that the theoretical value drops prematurely. 
The magnetization is moderately rounded for $h\agt1/3\,$, but does not 
have the plateau seen in experiment\cite{nojiri88}. 

\acknowledgments 

We are grateful to Prof. H. Shiba, Dr. U. Schotte and 
Prof. A. V. Chubukov for discussions and correspondence. 
This research was supported by the Natural Sciences and Engineering 
Research Council of Canada.

%1
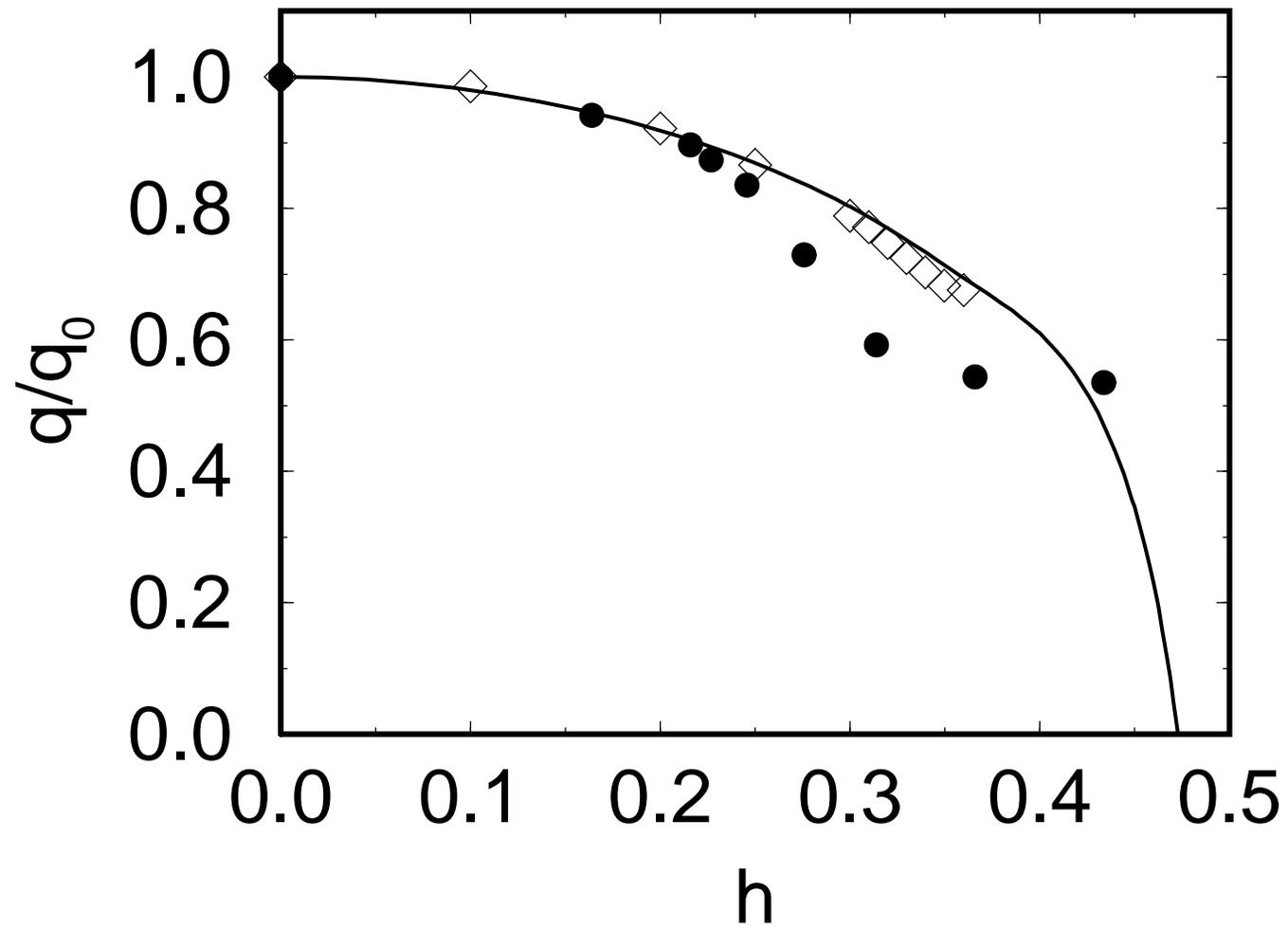
\begin{figure}
\caption{Dependence of the reduced wavenumber $q/q_0$ on the reduced 
field $h$. The solid circles are the experimental results of Ref. 20. 
Both the classical result (solid line, from Ref. 24) 
and the linear spin--wave result (diamonds) were found 
for in--plane spins and isotropic exchange $(\eta=0)$. 
}
\end{figure}

%2
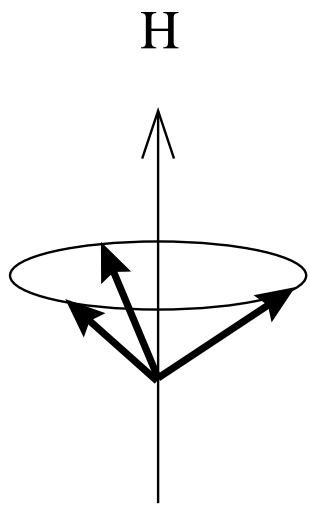
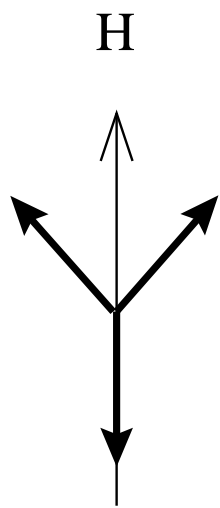
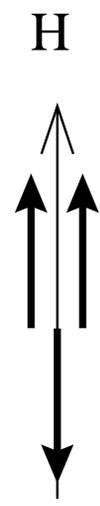
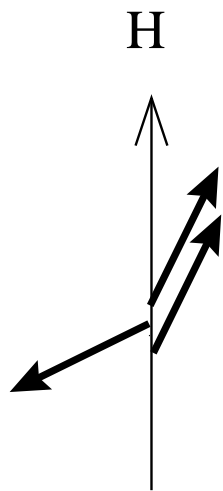
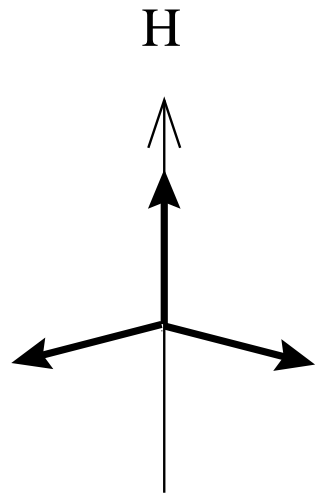
\begin{figure}
\caption{The five states, the umbrella state (a) and the four coplanar 
states (b) to (e), used in the examination of the phenomenological 
energy $E_{\rm fluct}\,$. 
} 
\end{figure}

%3
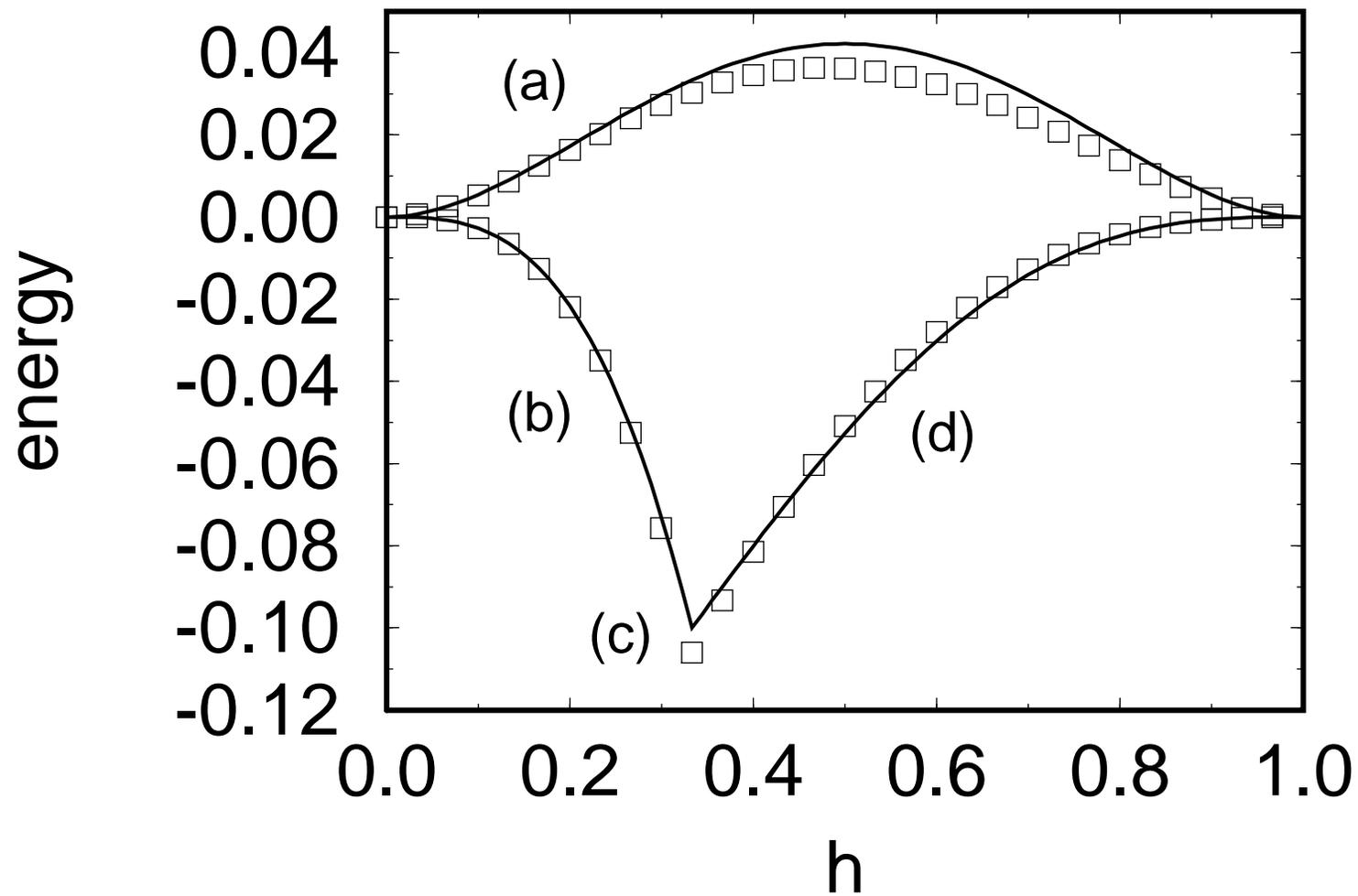
\begin{figure}
\caption{Comparison of the linear spin--wave result $E_2$ (squares) and the 
phenomenological energy $E_{\rm fluct}$ (lines, with $J_2=0.13\,$K, to first 
order in $J_2$) for the five commensurate CsCuCl$_3$ states corresponding 
to the states of Figure 2. 
The figure gives the energies $E_{\rm a}-E_{\rm e}$, {\it etc}, of states 
(a) to (d) relative to that of state (e); 
the energy of state (b) is plotted only for $h<1/3$, that of (c) only for 
$h=1/3$, and that of (d) only for $h>1/3$. 
}
\end{figure}

%4
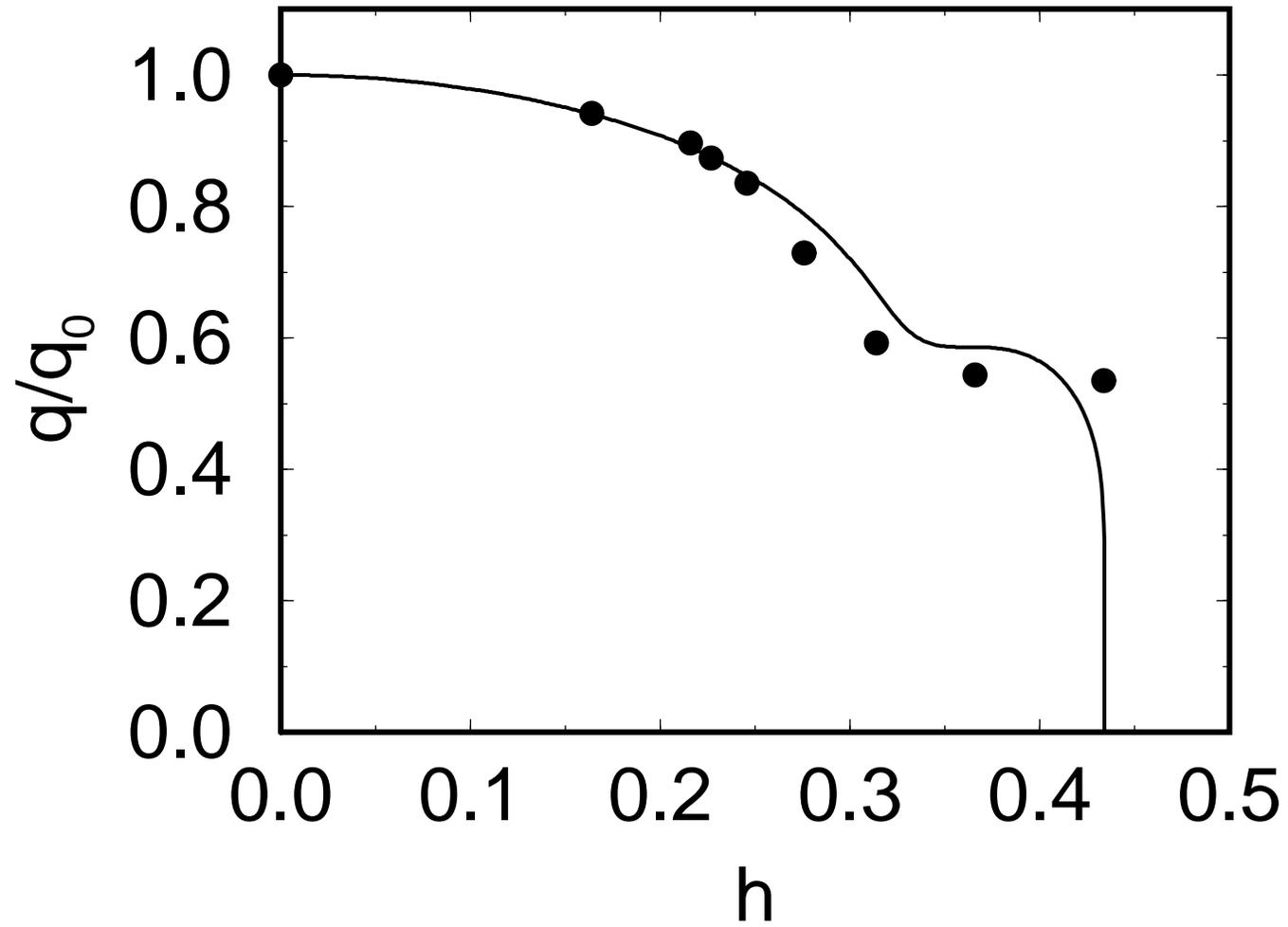
\begin{figure}
\caption{Dependence of the reduced wavenumber $q/q_0$ on the reduced field $h$. 
The circles are the experimental results of Ref. 20; 
the structure is unknown beyond $h\approx0.44\,$. 
The line gives the theoretical results based on a phenomenological 
treatment of quantum fluctuations ($J_2=0.13\,$K), for weak easy--plane 
anisotropy ($\eta=8.6\times10^{-3}$); 
the commensurate state is stable for $h>0.44\,$. 
} 
\end{figure}

\end{document}